\documentclass{article}

\usepackage{amsfonts,amsmath,amssymb,latexsym}

\usepackage{bbm} 

\def\real{\mathbb{R}}

\def\mat1{\mathbb{I}}

\newcommand { \Iletter}[1] {I\kern-0.10em #1 }

\def\bit{\begin{itemize}}
\def\eit{\end{itemize}}
\def\ben{\begin{enumerate}}
\def\een{\end{enumerate}}
\def\bde{\begin{description}}
\def\ede{\end{description}}

\def\bar{\begin{array}}
\def\ear{\end{array}}
\def\beq{\begin{equation}}
\def\eeq{\end{equation}}
\def\bfi{\begin{figure}[hbt] \begin{center}}
\def\efi{\end{center} \end{figure}}

\def\bce{\begin{center}}
\def\ece{\end{center}}
\newcommand{\proof}{{\bf Proof. }}
\newcommand{\cqfd}{\hfill $\Box$}

\newtheorem {rema} {Remark} [section]

\newtheorem {propo} {Proposition}

\title{Shannon Entropy Reinterpreted}
\author{L. Truffet \\
IMT-Atlantque \\
Dpt. Automatique-Productique-Informatique \\
La Chantrerie, 4 rue
A. Kastler, BP 20722, Nantes Cedex 3, FRANCE \\
email: laurent.truffet@imt-atlantique.fr \\
url: http://www.emn.fr/truffet}

\begin{document}

\maketitle
\begin{abstract}
In this paper we remark that Shannon entropy can be expressed as a
function of the self-information (i.e. the logarithm) and the inverse
of the Lambert $W$ function. It means that we consider that Shannon
entropy has the trace form: $-k \sum_{i} W^{-1} \circ
\mathsf{ln}(p_{i})$. Based on this remark we define a generalized
entropy which has as a limit the Shannon entropy. In order to
facilitate the reasoning this generalized entropy is obtained by a
one-parameter deformation of the logarithmic function.

Introducing a new concept of independence of two systems the Shannon
additivity is replaced by a non-commutative and non-associative law
which limit is the usual addition. The main properties associated with
the generalized entropy are established, particularly those
corresponding to statistical ensembles. The Boltzmann-Gibbs statistics
is recovered as a limit. The connection with thermodynamics is also
studied. We also provide a guideline for systematically defining a
deformed algebra which limit is the classical linear algebra. As 
an illustrative example we study a generalized
entropy based on Tsallis self-information. We point out possible 
connections between deformed algebra and fuzzy logics. Finally, noticing 
that the new concept of independence is based on t-norm the one-parameter 
deformation of the logarithm is interpreted as an additive generator of 
t-norms.
\end{abstract}


\section{Introduction}
\label{secIntro}

This paper is mainly motivated by three facts.

\ben
\item Within several scientific communities generalizations of the Shannon 
entropy have been developed. Such developements are motivated 
by the fact that apparently Boltzmann-Gibbs statistics fails 
to explain some observed results of physical systems with e.g. long-range 
interaction, long-time memory, non-Markovian 
systems, economic systems (see e.g. \cite[Section I]{kn:Tsallis95}, 
\cite{kn:Tribeche15} and references therein). 

\item When we want to establish an equilibrium thermal statistics which 
generalizes the Boltzmann-Gibbs one we optimize a generalized entropy 
under some constraints. For the microcanonical ensemble and internal 
energy most of the constraints are of the forms:

\begin{equation}
\label{eqU1}
\sum_{i} u_{1}(p_{i}) = Q,
\end{equation}
and 
\begin{equation}
\label{eqU2}
\sum_{i} u_{2}(p_{i}) \; \varepsilon_{i} = U,
\end{equation}
respectively. Where $Q \in \real$, 
$p_{i}$ denotes the probability of the $i$th-microstate. $u_{1}$ and 
$u_{2}$ are functions of the $p_{i}$'s. Typical choice for $u_{1}$ and $u_{2}$ is 
$p_{i}^{q}$ for some $q \in \real$ (in such case 
it is sometimes called {\em effective probability} \cite{kn:Wang01}). 
Another possible choice for $u_{1}, u_{2}$ is $\frac{p_{i}^{q}}{\sum_{j}p_{j}^{q}}$ 
(in such case it is called {\em escort probability} \cite{kn:Tsallisetal98}). 

The choice of the functions $u_{1}$ and $u_{2}$ seems not to be based on a clear 
procedure. 

\item The linear algebra is associated with the 
Shannon entropy. It is natural to ask which algebraic 
structure is associated with a generalized entropy. Once 
again it appears that there is no straightforward 
procedure which allows to associate a generalized 
entropy with a 'deformed algebra' (see among others \cite{kn:Kalo05}, 
\cite{kn:Nivanenetal03}, \cite{kn:Borges04}). 

\een

There exist many different ways to define a new entropy which 
generalizes the Shannon entropy:

\bit 
\item By imposing an entropy of the form (see e.g. \cite{kn:Renyi61}):
\[
G (\sum_{i} g(p_{i}))
\]
for some functions $G$ and $g$. 

\item By imposing a trace form of the entropy, i.e.:
\[
\sum_{i} g(p_{i})
\]
where $g$ can have the following form
\[
g(p_{i}):= p_{i}^{q} \; f_{\xi}(p_{i}),
\]
with $q \in \real$, $f_{\xi}(\cdot)$ playing the role of a deformed logarithm 
depending on the set of parameters $\xi$. 

The reader is refered to e.g. \cite[Table 1]{kn:Ilic2014}, 
\cite{kn:Oiko09} for more details on such entropies.

\item By noticing that the Shannon entropy can be defined as 
the following limit:
\[
\lim_{t \rightarrow -1} \frac{d}{dt} \sum_{i}p_{i}^{-t}
\]
and to generalize this definition by replacing the classical 
derivatives operator, i.e. $\frac{d}{dt}$, by Jackson or fractional 
derivatives (see e.g. \cite{kn:Ubriaco09}).

\item More recently, by imposing that for two sets of observable states 
$A$ and $B$ which are independent the entropy $S(A \times B)$ satisfies 
\[
S(A \times B)= S(A) \hat{+} S(B) 
:= S(A) + S(B) + \sum_{i,j} c_{i,j} \; S^{i}(A) \; S^{j}(B).
\]
The coefficients $c_{i,j}$ are such that $\hat{+}$ is commutative, 
associative and has a neutral element. For more details on the 
subject see e.g. \cite{kn:Tempesta2016b}. The composition law 
$\hat{+}$ can be seen as a 
generalization of Sugeno-Tsallis composition 
formula (see \cite{kn:Tsallis88}, \cite{kn:Sugenothese}): 
$S(A \times B)= S(A) + S(B) + (1-q) \; S(A) \;  S(B)$.

\eit

In this paper we propose to define new entropy 
based on a reinterpretation of the Shannon entropy which 
is explained hereafter.

Let $n \geq 1$ and $\Omega :=\{\omega_{1}, \ldots, \omega_{n} \}$ be 
the set of observable states of a given system $\mathcal{S}$ \footnote{Usually 
the number of states is denoted by $W$ instead of $n$ 
but here $W$ denotes the Lambert function.}. We associate 
with $\Omega$ the following application:

\begin{equation}
\bar{ll}
p^{\Omega}: & \Omega \longrightarrow [0,1] \\
\mbox{ } & \omega_{i} \longmapsto p^{\Omega}_{i}.  
\ear
\end{equation}
$p^{\Omega}_{i}$ is interpreted as a certain frequency of 
apparition of the observable state $\omega_{i}$. If there is no 
ambiguity with the context $p^{\Omega}$ (resp. $p^{\Omega}_{i}$) 
will be simply denoted $p$ (resp. $p_{i}$).

The Shannon entropy is defined on $\Omega$ as follows:

\begin{equation}
S^{\Omega}_{\mathsf{Sh}} := - k \; \sum_{i=1}^{n} p_{i} \; \mathsf{ln}(p_{i})
\end{equation}
where $\mathsf{ln}(\cdot)$ denotes the Neperian logarithm. 

If there is no ambiguity $S^{\Omega}_{\mathsf{Sh}}$ is simply 
denoted by $S_{\mathsf{Sh}}$.

We reformulate the Shannon entropy based on:
\bit 
\item The notion of self-information\footnote{In the litterature the self-information or surprisal is defined as a nonnegative quantity which is the 
opposite of the one used in this paper.}, i.e. the function:
\begin{equation}
\bar{ll}
I_{\mathsf{Sh}}: & [0,1] \longrightarrow [-\infty,0] \\
\mbox{ } & p_{i} \longmapsto \mathsf{ln}(p_{i}).
\ear
\end{equation}

\item And $W^{-1}$ the inverse of the Lambert W function, i.e.:
\begin{equation}
\bar{ll}
W^{-1}: & \real \longrightarrow \real \\
\mbox{ } & x \longmapsto x \; e^{x}.
\ear
\end{equation}
\eit

Shannon entropy is thus reformulated as the following trace 
form:
\begin{equation}
\label{Shannon2.1}
S_{\mathsf{Sh}} = - k \; \sum_{i=1}^{n} W^{-1} \circ I_{\mathsf{Sh}}(p_{i}) =-k \; 
\sum_{i=1}^{n}  I_{\mathsf{Sh}}(p_{i}) \;  e^{I_{\mathsf{Sh}}(p_{i})}.
\end{equation}

In the sequel the positive (Boltzmann) constant $k$ whose values depends 
on the particular units to be used is set for simplicity equal to $1$.

From the trace form (\ref{Shannon2.1}) we remark that to obtain 
a generalized entropy it is sufficient to generalize the notion of 
self-information $I_{\mathsf{Sh}}$. For seek of simplicity we assume 
that the generalized self-information only depends upon one 
real parameter, say $a$, and is denoted $I_{a}$. In other words 
we just have to deform the logarithmic function to obtain a 
deformed entropy which generalizes the Shannon entropy. Thus, we 
introduce the following function:

\begin{equation}
\label{eqSa}
\bar{ll}
S_{a}: & [0,1]^{n} \longrightarrow \real \\
\mbox{ } & p \longmapsto - \sum_{i=1}^{n} W^{-1} \circ I_{a}(p_{i}),
\ear
\end{equation}
with: 
\[
I_{a} :[0,1] \rightarrow [-\infty, 0].
\]

We require that there exists $V \subseteq \real$ such that 
for all $a \in V$ the function $S_{a}$ satisfies the first 
three Shannon-Khinchin axioms. It has been shown to be equivalent 
to assume (see e.g. \cite{kn:Hanel011}) that for all $a \in V$:

\bit
\item ({\bf A1}). $W^{-1} \circ I_{a}$ is convex on $[0,1]$ 
\item ({\bf A2}). $W^{-1} \circ I_{a}$ is continuous 
\item ({\bf A3}). $W^{-1} \circ I_{a}(0)=0$
\eit

The reformulation of Shannon entropy (\ref{Shannon2.1}) 
based on self-information suggests to introduce the following definitions. 

\begin{def}
\label{mainDef}
The main definitions of this paper are hereafter. 

\bit
\item $a$-deformed number. Let $x \in [0,1]$ let us define:
\begin{equation}
[x]_{a}:= e^{I_{a}(x)}.
\end{equation}
When $x$ is interpreted as a probability then we use the 
term $a$-probability.

\item $a$-deformed statistics. The $a$-expectation value $\langle O \rangle_{a}$ 
of an observable $O$ is defined by:

\begin{equation}
\langle O \rangle_{a}:= \sum_{i} [p_{i}]_{a} \; O_{i}
\end{equation}
where $O_{i}$ is the value of $O$ at state $i$, $[p_{i}]_{a}$ is 
an $a$-probability. 
Note that $\langle 1 \rangle_{a}$ corresponds to the particular case 
where $\forall i, O_{i}=1$.

\item $a$-independence. Let us consider two sets of observable states 
$\Omega=\{\omega_{1}, \ldots, \omega_{n} \}$ and $\Omega'=\{\omega_{1}', 
\ldots, \omega_{m}' \}$ of systems $\mathcal{S}$ and $\mathcal{S}'$, 
respectively. 
The two systems  $\mathcal{S}$ and $\mathcal{S}'$ are said to be 
$a$-independent if:
\begin{equation}
\label{eqa-independence}
\forall i=1, \ldots, n; \forall j=1, \ldots , m, \quad 
[p^{\Omega \times \Omega'}_{ij}]_{a} = [p^{\Omega}_{i}]_{a} \; [p^{\Omega'}_{j}]_{a}
\end{equation}
where $[p^{\Omega \times \Omega'}_{ij}]_{a}$, $[p^{\Omega}_{i}]_{a}$ and 
$[p^{\Omega'}_{j}]_{a}$ denote the $a$-probabilities of being 
in state $(\omega_{i}, \omega_{j}')$, $\omega_{i}$ and $\omega_{j}'$, 
respectively. 

\eit 

\end{def}

\begin{rema}
\label{remark1}
Note that an $a$-probability vector $([p_{1}]_{a}, \ldots , [p_{n}]_{a})$ 
is a nonnegative vector which is not a probability vector in general, i.e.:
\[
\sum_{i=1}^{n} [p_{i}]_{a}  \neq 1.
\]
And thus, an $a$-expectation is not the mean value associated with a
random variable defined on $\Omega=\{\omega_{1}, \ldots, \omega_{n}
\}$. Maybe one of the simplest possible choices for $I_{a}(x)$ is $a \;
\mathsf{ln}(x)$ with $a \geq 0$. This choice leads to an $a$-probability 
vector of the well-known form: $(p_{1}^{a}, \ldots, p_{n}^{a})$. And we refer to e.g. 
\cite[Sections 1 and 2]{kn:Wang01} who discussed the interpretation of 
an effective probability in the context of incomplete statistics. 
Since $I_{a} \leq 0$ we have $e^{I_{a}} \leq 1$ and this discussion could also 
be applied to the more general case of an $a$-probability vector.
\end{rema}

The results and the organization of the paper are as follows. 
In Section~\ref{secMR} main results of the paper are presented. 
We follow the approach developed in \cite{kn:Jayne57}. We 
assume some smoothness of $I_{a}$ (i.e. once derivability) 
that $I_{a}$ has an inverse denoted $I_{a}^{-1}$ and 
that there exists $b \in \overline{V}$\footnote{$\overline{V}$ denotes 
the closure of the set $V$} such that the $\rightarrow b$ limit 
of $I_{a}$ is $\mathsf{ln}$ and the $\rightarrow b$ limit of 
$I_{a}^{-1}$ is the exponential function $e$. Thus, a natural choice 
for $u_{1}$, $u_{2}$ in equations (\ref{eqU1}) and  (\ref{eqU2}) is 
$e^{I_{a}(p_{i})}$ which $\rightarrow b$ limit is $p_{i}$. In 
subsection~\ref{subMicrocanonical} we prove 
that $S_{a}(p_{1},\ldots, p_{n})$ is maximal when 
$\forall i=1, \ldots n: \quad p_{i} = I_{a}^{-1} \circ 
\mathsf{ln}\left(\frac{\langle 1 \rangle_{a}}{n} \right)$. For two systems assumed 
to be $a$-independent the Shannon additivity is replaced by 
a non-commutative and non-associative law of the form:
\[
(x,y) \mapsto \rho \; x + \theta \; y
\]
which $\rightarrow b$ limit is the usual addition. This is the 
result of subsection~\ref{subAdd}. Properties corresponding to 
statistical ensembles are studied in subsection~\ref{subCanonical}. And 
in subsection~\ref{subTempEner} a connection with thermodynamics is 
established. Finally, in subsection~\ref{subGuideline} we give 
a way to systematically define deformed addition $\stackrel{a}{\oplus}$ 
by requiring that:
\[
e^{I_{a}(x\stackrel{a}{\oplus}y)} = e^{I_{a}(x)} + 
e^{I_{a}(y)} 
\]
which leads to
\begin{equation}
\label{eqGenAdd}
x \stackrel{a}{\oplus} y := I_{a}^{-1} \circ \mathsf{ln} \left(e^{I_{a}(x)} + 
e^{I_{a}(y)} \right) 
\end{equation}
and define a multiplication $\stackrel{a}{\otimes}$ by requiring that:
\[
e^{I_{a}(x\stackrel{a}{\otimes}y)} = e^{I_{a}(x)} \times e^{I_{a}(y)} 
\]
which leads to
\begin{equation}
\label{eqGenMult}
x \stackrel{a}{\otimes} y := I_{a}^{-1} \circ \mathsf{ln}\left(e^{I_{a}(x)} \times 
e^{I_{a}(y)} \right).
\end{equation}
This way of defining operations seems to be known at least since 
\cite{kn:Kaniadakis02}. Noticing that $I_{a}^{-1} \circ \mathsf{ln}$ is 
the inverse function of $e \circ I_{a}$ the deformed multiplication distributes 
over the deformed addition. Moreover, these operations are associative and 
commutative. These operations are particular cases of  
pseudo-addition and pseudo-multiplication which appear in the context of 
pseudo-analysis (see e.g. \cite{kn:MesiarPap99} and references therein).

In Section~\ref{secEx} we illustrate the results of Section~\ref{secMR}. The 
deformed self-information we use is the log-exp transform of the 
Schweizer-Sklar-Tsallis logarithm also known as Frank generator 
\cite{kn:Nelsen06}. 

In Section~\ref{secConcl} we conclude by recalling the main features
of the paper. We point out a possible connection between the
deformed algebra developed in Section~\ref{secEx} and t-norm fuzzy
logics. Finally, as suggested by one of the reviewers we suggest a way 
of inverting our approach: i.e. given a non-standard notion of independence 
we show that it seems possible to derive a deformed self-information 
and then a deformed entropy.

\section{Main results}
\label{secMR}

In this Section we follow the methodology developed 
in e.g. \cite{kn:CuTs91} and \cite{kn:Tsallis88}. 

Let $V \subseteq \real$ and for all $a \in V$ let $S_{a}$ be 
the function defined by (\ref{eqSa}). Because $I_{a}$ is $\leq 0$ 
and the assumption ({\bf A1}) $S_{a}$ is clearly a nonnegative 
Schur-concave function.

In the sequel we assume that $I_{a}$ is a smooth function which acts
as a deformed logarithm. Thus, $I_{a}$ satisfies the following
assumptions:

\bit
\item ({\bf I1}). $I_{a}$ is invertible 
on $I_{a}([0,1])$. It's inverse is denoted $I_{a}^{-1}$. 
\item ({\bf I2}). $\exists b \in \overline{V}$, $\forall x \in [0,1] 
\lim_{a \rightarrow b}I_{a}(x) = I_{\mathsf{Sh}}(x), \quad \lim_{a \rightarrow b} I_{a}^{-1}(x) = I_{\mathsf{Sh}}^{-1}(x)$. 
\item ({\bf I3}). Smoothness: $I_{a}$ is derivable on $(0,1]$. 
And $I_{a}' \neq 0$ on $(0,1)$.
\eit

\subsection{Microcanonical ensemble}
\label{subMicrocanonical}

We are looking for a candidate of the following optimization 
problem:
\begin{equation}
\max_{p} \left\{S_{a} : \sum_{i=1}^{n}[p_{i}]_{a}= \langle 1 \rangle_{a} \right\}.
\end{equation}
Where $\langle 1 \rangle_{a}$ is assumed to be independent of the $p_{i}$'s.

Let us introduce the $\lambda_{a}$ Lagrange parameter and define 
the function:
\begin{equation}
\bar{ll}
\phi(p,\lambda_{a}) & := S_{a} + \lambda_{a} \; (\sum_{i=1}^{n}[p_{i}]_{a} - \langle 1 \rangle_{a}) \\
\mbox{ }           & =  - \sum_{i=1}^{n} I_{a}(p_{i}) \; e^{I_{a}(p_{i})} + 
\lambda_{a} \; (\sum_{i=1}^{n}  e^{I_{a}(p_{i})} - \langle 1 \rangle_{a}).
\ear
\end{equation}
Because $I_{a}$ is derivable (see ({\bf I3})) we have:

\[
\frac{\partial \phi}{\partial p_{i}} = I_{a}'(p_{i}) \; e^{I_{a}(p_{i})} \; 
(-1-I_{a}(p_{i}) + \lambda_{a})
\]
and 
\[
\frac{\partial \phi}{\partial \lambda_{a}}= \sum_{i=1}^{n}[p_{i}]_{a} - \langle 1 \rangle_{a}.
\]
Imposing $\forall i=1, \ldots , n$: $\frac{\partial \phi}{\partial p_{i}}=0$ 
and $\frac{\partial \phi}{\partial \lambda_{a}}=0$ one obtains because 
$I_{a}' \neq 0$ (see ({\bf I3})):
\begin{equation}
\label{eqMicro0}
\bar{lll}
\forall i, & I_{a}(p_{i}) & = -1 + \lambda_{a} \\
\mbox{}    & -1 + \lambda_{a} & = \mathsf{ln} \left(\frac{\langle 1 \rangle_{a}}{n} \right).
\ear
\end{equation}
Thus, 
\begin{equation}
\label{eqMicro}
\forall i, p_{i} = I_{a}^{-1} \circ \mathsf{ln} \left(\frac{\langle 1 \rangle_{a}}{n} \right).
\end{equation}
Under this condition on the $p_{i}$'s we have
\begin{equation}
\label{eqMicroSa}
S_{a}= \langle 1 \rangle_{a} \; \mathsf{ln}(n).
\end{equation}

The $p_{i}$'s correspond to a kind of deformed uniform law on $\Omega$. 
By assumption ({\bf I2}) on $I_{a}$ the $\rightarrow b$ limit 
of $\langle1\rangle_{a}$ is $1$ then the $\rightarrow b$ limit 
of the $p_{i}$'s corresponds to the equiprobability case. And it is immediately 
verified that the Boltzmann expression is recovered 
as the $\rightarrow b$ limit of $S_{a}$, i.e.: 
\[
S_{b}= \mathsf{ln}(n).
\]

\subsection{Additivity}
\label{subAdd}
Let us consider two sets of observable states 
$\Omega=\{\omega_{1}, \ldots, \omega_{n} \}$ and $\Omega'=\{\omega_{1}', 
\ldots, \omega_{m}' \}$ of systems $\mathcal{S}$ and $\mathcal{S}'$, 
respectively. 

We have the following result.

\begin{propo}
\label{propAdditivite}
If the systems $\mathcal{S}$ and $\mathcal{S}'$ are $a$-independent 
then
\[
S^{\Omega \times \Omega'}_{a} = \langle 1 \rangle_{a}' \; S^{\Omega}_{a} + \langle 1 \rangle_{a} \; S^{\Omega'}_{a},
\]
where $\sum_{i=1}^{n}[p^{\Omega}_{i}]_{a} =\langle 1 \rangle_{a}$ and 
$\sum_{j=1}^{m}[p^{\Omega'}_{j}]_{a} =\langle 1 \rangle_{a}'$.

\end{propo}
\proof By definition of $S^{\Omega \times \Omega'}_{a}$ we have:
\[
S^{\Omega \times \Omega'}_{a} = -\sum_{i=1}^{n} \sum_{j=1}^{m} 
[p^{\Omega \times \Omega'}_{ij}]_{a} \; I_{a}(p^{\Omega \times \Omega'}_{ij}),
\]

Now, assuming the $a$-independence of the two systems 
$\mathcal{S}$ and $\mathcal{S}'$, by definition of $[\cdot]_{a}$ and 
$I_{a}$ we have: $I_{a}(p^{\Omega \times \Omega'}_{ij}) = I_{a}(p^{\Omega}_{i}) + 
I_{a}(p^{\Omega'}_{j})$ and thus:
\[
S^{\Omega \times \Omega'}_{a} = -\sum_{i=1}^{n} \sum_{j=1}^{m} 
[p^{\Omega}_{i}]_{a} \; [p^{\Omega'}_{j}]_{a} \; (I_{a}(p^{\Omega}_{i}) + 
I_{a}(p^{\Omega'}_{j}))
\]
By definition of $S^{\Omega}_{a}$ and $S^{\Omega'}_{a}$ and recalling 
that: $\sum_{i=1}^{n}[p^{\Omega}_{i}]_{a} =\langle 1 \rangle_{a}$ and 
$\sum_{j=1}^{m}[p^{\Omega'}_{j}]_{a} =\langle 1 \rangle_{a}'$ the result is proved. \cqfd

The composition law $(x,y) \mapsto \langle 1 \rangle_{a}' \; x + \langle 1 
\rangle_{a} \; y$ 
is neither commutative nor associative in 
general except in the case where $\langle 1 \rangle_{a}= \langle 1 \rangle_{a}'=1$. 

Taking the $\rightarrow b$ limit we retrieve the Shannon entropy 
additivity. The $b$-independence of the systems 
$\mathcal{S}$ and $\mathcal{S}'$ reduces to the classical 
probabilistic independence of $\mathcal{S}$ and $\mathcal{S}'$.

\subsection{Canonical ensemble}
\label{subCanonical}
We want to find a candidate to the following optimization problem:

\begin{equation}
\max_{p} \left\{S_{a} : \sum_{i=1}^{n}[p_{i}]_{a}= \langle 1 \rangle_{a}, 
\quad \sum_{i=1}^{n}[p_{i}]_{a} \varepsilon_{i}=U_{a} \right\}.
\end{equation}

To do this let us introduce the Lagrange parameters $\lambda_{a}$ and 
$\mu_{a}$ and define the following function:

\begin{equation}
\psi(p,\lambda_{a}, \mu_{a}):= S_{a} + \lambda_{a} (\sum_{i=1}^{n}[p_{i}]_{a}- 
\langle1\rangle_{a}) + \mu_{a} (\sum_{i=1}^{n}[p_{i}]_{a} \varepsilon_{i} -U_{a}).
\end{equation}

Imposing $\forall i=1, \ldots , n$: $\frac{\partial \psi}{\partial p_{i}}=0$, 
$\frac{\partial \psi}{\partial \lambda_{a}}=0$ and 
$\frac{\partial \psi}{\partial \mu_{a}}=0$ one obtains:

\begin{equation}
\label{eqCanon0}
\bar{llll}
(a) & \forall i, & I_{a}(p_{i}) & = -1 + \lambda_{a} + \mu_{a} \; \varepsilon_{i} \\
(b) & \mbox{}    & e^{-1 + \lambda_{a}} \; Z(\mu_{a}) & = \langle 1 \rangle_{a} \\
(c) & \mbox{}    & e^{-1 + \lambda_{a}} \; \frac{\partial Z}{\partial \mu_{a}} & = 
U_{a},
\ear
\end{equation}
where
\begin{equation}
\label{eqCanon01}
Z(\mu_{a}):= \sum_{i=1}^{n} e^{\mu_{a} \varepsilon_{i}}.
\end{equation}
Thus, one deduces that:
\begin{equation}
\label{eqCanon}
\forall i, \; I_{a}(p_{i}) = -\mathsf{ln}(Z(\mu_{a})) + \mathsf{ln}(\langle 1 
\rangle_{a}) + \mu_{a} \; \varepsilon_{i}.
\end{equation}
Or, equivalently that:
\begin{equation}
\label{eqCanon2}
\forall i, \; p_{i} = I_{a}^{-1} \circ \mathsf{ln} \left(\frac{\langle 1 \rangle_{a} \; e^{\mu_{a} \varepsilon_{i}}}{Z(\mu_{a})} \right).
\end{equation}

Integrating (\ref{eqCanon0}, (b) and (c)) one obtains:
\begin{equation}
\label{eqCanon3}
Z(\mu_{a}) = e^{\mu_{a} \frac{U_{a}}{\langle 1 \rangle_{a}}}.
\end{equation}
Thus, one derives a second expression for the $p_{i}$'s as follows:

\begin{equation}
\label{eqCanon4}
\forall i, \; p_{i}(\varepsilon_{i}, U_{a}) = 
I_{a}^{-1} \circ 
\mathsf{ln} \left(\langle 1 \rangle_{a} \; e^{\mu_{a} (\varepsilon_{i}-\frac{U_{a}}{\langle 1 \rangle_{a}})} \right)
\end{equation}
which is invariant by the transformation $(\varepsilon_{i}, U_{a}) 
\rightarrow (\varepsilon_{i} + c, U_{a} + \langle 1 \rangle_{a} \; c)$ for 
$c \in \real$.

By assumption ({\bf I2}) on $I_{a}$  it 
is immediately verified that 
in the $\rightarrow b$ limit one retrieves the Gibbs 
measure, i.e.:

\[
\forall i, \; p_{i} = \frac{e^{\mu_{b} \varepsilon_{i}}}{Z(\mu_{b})}=e^{\mu_{b} 
(\varepsilon_{i} - U_{b})}.
\]
Which is invariant by the transformation $(\varepsilon_{i}, U_{b}) 
\rightarrow (\varepsilon_{i} + c, U_{a} + c)$.

\subsection{Generalized temperature and free energy}
\label{subTempEner}
From the expression (\ref{eqCanon}) we derive the 
generalized temperature as follows.

\[
\bar{ll}
S_{a} & = - \sum_{i=1}^{n} (-\mathsf{ln}(Z(\mu_{a})) + \mathsf{ln}(\langle 1 \rangle_{a}) + 
\mu_{a} \; \varepsilon_{i}) \; e^{-\mathsf{ln}(Z(\mu_{a})) + \mathsf{ln}(\langle 1 \rangle_{a}) + \mu_{a} \varepsilon_{i})} \\
\mbox{ } &= - \sum_{i=1}^{n} (-\mathsf{ln}(Z(\mu_{a})) + \mathsf{ln}(\langle 1 \rangle_{a}))
 \; e^{-\mathsf{ln}(Z(\mu_{a})) + \mathsf{ln}(\langle 1 \rangle_{a}) + \mu_{a} \varepsilon_{i})}  \\
\mbox{} & - \sum_{i=1}^{n} \mu_{a} \; \varepsilon_{i} \; 
e^{-\mathsf{ln}(Z(\mu_{a})) + \mathsf{ln}(\langle 1 \rangle_{a}) + \mu_{a} \varepsilon_{i})}.
\ear
\]
Based on (\ref{eqCanon0}), (\ref{eqCanon01}) and 
(\ref{eqCanon}) we have: 
\[
- \sum_{i=1}^{n} (-\mathsf{ln}(Z(\mu_{a})) + \mathsf{ln}(\langle1\rangle_{a}))
e^{-\mathsf{ln}(Z(\mu_{a})) + \mathsf{ln}(\langle1\rangle_{a}) + \mu_{a} \varepsilon_{i})} 
= \langle1\rangle_{a} (\mathsf{ln}(Z(\mu_{a})) - \mathsf{ln}(\langle1\rangle_{a})),
\]
and
\[ 
\sum_{i=1}^{n} \mu_{a} \; \varepsilon_{i} \; e^{-\mathsf{ln}(Z(\mu_{a})) + \mathsf{ln}(\langle1\rangle_{a}) + \mu_{a} \varepsilon_{i})} = \mu_{a} \; U_{a}.
\]
Thus, 
\begin{equation}
S_{a}= \langle1\rangle_{a} \; (\mathsf{ln}(Z(\mu_{a})) - \mathsf{ln}(\langle 1 \rangle_{a})) - \mu_{a} \; U_{a}. 
\end{equation}

So, defining $\frac{1}{T_a} := \frac{\partial S_{a}}{\partial U_a}$ 
we have:

\begin{equation}
\frac{1}{T_{a}} = - \mu_{a} .
\end{equation}

Defining the free energy as $F_{a}:= U_{a} - T_{a} S_{a}$, one obtains that:

\begin{equation}
F_{a} = \frac{\langle 1 \rangle_{a}}{\mu_{a}} \; (\mathsf{ln}(Z(\mu_{a})) -\mathsf{ln}(\langle 1 \rangle_{a}),
\end{equation}
which recovers for $a \rightarrow b$ the classical 
limit, namely $F_{b}= \frac{1}{\mu_{b}} \; \mathsf{ln}(Z(\mu_{b}))$.

\subsection{A guideline to obtaining deformed algebra and calculus}
\label{subGuideline}
Let us consider a function $I_{a}$ satisfying assumptions ({\bf
  I1})-({\bf I3}). Moreover, for seek of simplicity let us assume that
$I_a(0)=-\infty$ and $I_{a}(1)=0$.

Let us recall that deformed addition (see 
(\ref{eqGenAdd})) and multiplication (see (\ref{eqGenMult})) 
are defined as follows:

\[
x \stackrel{a}{\oplus} y := I_{a}^{-1} \circ \mathsf{ln} \left(e^{I_{a}(x)} + 
e^{I_{a}(y)} \right) 
\]
and
\[
x \stackrel{a}{\otimes} y := I_{a}^{-1} \circ \mathsf{ln} \left(e^{I_{a}(x)} \times 
e^{I_{a}(y)} \right) = I_{a}^{-1}(I_{a}(x) + I_{a}(y)).
\]
By their structure these operations are commutative and associative. $0$ 
(resp. $1$) is the neutral element for $\stackrel{a}{\oplus}$ (resp. 
$\stackrel{a}{\otimes}$) and $\stackrel{a}{\otimes}$ distributes over 
$\stackrel{a}{\oplus}$. 

By assumptions on $I_{a}$ it is clear that the $\rightarrow b$ limit 
of $\stackrel{a}{\oplus}$ and $\stackrel{a}{\otimes}$ are the 
usual addition and multiplication, respectively.

Taking care about the use of logarithmic function it is possible to
define a deformed substraction and a deformed division as follows:

\begin{equation}
\label{eqGenMoins}
x \stackrel{a}{\ominus} y := I_{a}^{-1} \circ \mathsf{ln} \left(e^{I_{a}(x)} - 
e^{I_{a}(y)} \right) 
\end{equation}
and 
\begin{equation}
\label{eqGenDiv}
x \stackrel{a}{\oslash} y := I_{a}^{-1} \circ \mathsf{ln} \left(\frac{e^{I_{a}(x)}}{
e^{I_{a}(y)}} \right).
\end{equation}
Their $\rightarrow b$ limits correspond to the usual substraction and 
division.

For a small element $dx$ one can define:

\[
(x+dx)\stackrel{a}{\ominus} x = I_{a}^{-1} \circ \mathsf{ln}(I_{a}'(x)
 e^{I_{a}(x)} dx)
\]
and
\[
f(x+dx)\stackrel{a}{\ominus} f(x)= I_{a}^{-1} \circ \mathsf{ln}(I_{a}'(f(x)) 
e^{I_{a}(f(x))} f'(x) dx).
\]

This leads to propose as deformed derivative the following limit:

\begin{equation}
\label{eqGenDerivee}
\bar{ll}
D_{a}f(x) &:= \lim_{dx \rightarrow 0}(f(x+dx)\stackrel{a}{\ominus} f(x)) \stackrel{a}{\oslash} ((x+dx)\stackrel{a}{\ominus} x) \\
\mbox{}  &= I_{a}^{-1} \circ 
\mathsf{ln} \left(\frac{(I_{a}'(f(x)) e^{I_{a}(f(x))}}{I_{a}'(x) e^{I_{a}(x)}} \;f'(x) \right)
\ear
\end{equation}

If \\
({\bf I4}): $\lim_{a \rightarrow b} I_{a}'(x) = \frac{1}{x}$ \\

then the $\rightarrow b$ limit of $D_{a}f$ is $f'$.

By noticing that:

\[
e^{I_{a}(D_{a}f(x))} = \frac{I_{a}'(f(x)) e^{I_{a}(f(x))}}{I_{a}'(x) e^{I_{a}(x)}} \; f'(x),
\]

\[
D_{a}f(x) \stackrel{a}{\otimes} g(x)=I_{a}^{-1} \circ \mathsf{ln} \left(e^{I_{a}(D_{a}f(x))} \times e^{I_{a}(g(x))} \right),
\]

\[
D_{a}f(x) \stackrel{a}{\otimes} g(x) \stackrel{a}{\oplus} f(x) 
\stackrel{a}{\otimes} D_{a}g(x) = 
I_{a}^{-1} \circ \mathsf{ln} \left(\frac{(I_{a}(f(x)) + I_{a}(g(x)))'}{I_{a}'(x) e^{I_{a}(x)}} \; e^{I_{a}(f(x)) + I_{a}(g(x))} \right)
\]
and that:

\[
\bar{ll}
D_{a}(f(x) \stackrel{a}{\otimes} g(x)) & = I_{a}^{-1} \circ \mathsf{ln} \left(
\frac{I_{a}'(f(x)\stackrel{a}{\otimes} g(x)) e^{I_{a}(f(x) \stackrel{a}{\otimes} g(x))}}
{I_{a}'(x) e^{I_{a}(x)}} \; (f(x)\stackrel{a}{\otimes} g(x))' \right) \\
\mbox{} & = I_{a}^{-1} \circ 
\mathsf{ln} \left(\frac{(I_{a}(f(x)\stackrel{a}{\otimes} g(x)))'}{I_{a}'(x) e^{I_{a}(x)}} \; e^{I_{a}(f(x) + g(x))} \right) \\
\mbox{} & = I_{a}^{-1} \circ 
\mathsf{ln} \left(\frac{(I_{a}(f(x)) + I_{a}(g(x)))'}{I_{a}'(x) e^{I_{a}(x)}} \; 
e^{I_{a}(f(x) + g(x))} \right)
\ear
\]
the Leibniz rule is verified, i.e.:

\[
D_{a}(f(x) \stackrel{a}{\otimes} g(x)) = D_{a}f(x) \stackrel{a}{\otimes} g(x) \stackrel{a}{\oplus} f(x) \stackrel{a}{\otimes} D_{a}g(x).
\]

We define a deformed integral $\int^{a}$ by requiring 
that $D_{a} \int^{a}f =f$ and we obtain:

\begin{equation}
\label{eqGenIntegral}
\int^{a} f = I_{a}^{-1} \circ \mathsf{ln} \left(\int I_{a}'(x) e^{I_{a}(x)} e^{I_{a}(f(x))}dx \right)
\end{equation}
where $\int$ denotes the Lebesgue integral and it is assumed that 
$x \mapsto I_{a}'(x) e^{I_{a}(x)} e^{I_{a}(f(x))}$ is Lebesgue integrable. 

We also remark 
that under the assumptions ({\bf I1})-({\bf I4}) and if \\
({\bf I5}): $\exists h \geq 0$ Lebesgue integrable s.t.:
\[
\forall a, \forall x: |I_{a}'(x) e^{I_{a}(x)} e^{I_{a}(f(x))}| \leq h(x)
\]
the $\rightarrow b$ 
limit of $\int^{a}$ is $\int$ (see e.g.\cite[p. 94]{kn:Royden88}).

From definition (\ref{eqGenIntegral}) one easily verifies that: 

\[
\int^{a} f \stackrel{a}{\oplus} g = \int^{a} f \stackrel{a}{\oplus}  \int^{a} g.
\]

\section{An illustrative example: generalized entropy based 
on Schweizer-Sklar-Tsallis self-information}
\label{secEx}

As an example, let $ a \in \real_{+} \setminus \{0\}$. And let the 
Tsallis self-information defined by (see e.g. \cite{kn:Tsallis88}):
\begin{equation}
\label{eqSelfTs}
\forall x \in [0,1], \; I^{\mathsf{Ts}}_{a} (x):= \frac{x^{a}-1}{a}.
\end{equation}

From Tsallis self-information let us define the following function:

\begin{equation}
\label{eqGSelfTs}
\forall x \in [0,1], \; I^{\mathsf{GTs}}_{a} (x):= \mathsf{ln} \circ I^{\mathsf{Ts}}_{a} \circ e^{x} - \mathsf{ln} \circ I^{\mathsf{Ts}}_{a} \circ e^{1}
\end{equation}
which could be rewritten as follows:

\begin{equation}
\label{eqGTs}
\forall x \in [0,1], \; I^{\mathsf{GTs}}_{a} (x) = \mathsf{ln} \left(\frac{e^{ax}-1}{e^{a}-1} \right).
\end{equation}
This function is also known as the Frank generator.

Let $V:=(0,\mathsf{ln}(1+e^{3})]$, 
thus $\overline{V}=[0,\mathsf{ln}(1+e^{3})]$. For all $a \in V$, 
it is clear that $W^{-1} \circ I^{\mathsf{GTs}}_{a}$ is continuous and 
Assumption ({\bf A2}) is verified. Adopting the convention 
that $\mathsf{ln}(0)=-\infty$, Assumption ({\bf A3}) is also 
verified. The convexity of $W^{-1} \circ I^{\mathsf{GTs}}_{a}$ (i.e. Assumption 
({\bf A1})) is proved in subsection~\ref{subSchur2}.

We remark that  $I^{\mathsf{GTs}}_{a}$ is invertible and 
its inverse is:
\begin{equation}
\label{InvGTs}
I^{- \mathsf{GTs}}_{a}(x)= \frac{1}{a} \mathsf{ln}(1 + (e^{a}-1) \; e^{x}).
\end{equation}
Thus, ({\bf I1}) is verified.

We note that for $b=0$ ({\bf I2}) is verified. 
The derivative of $I^{\mathsf{GTs}}_{a}$ is 
\[
(I^{\mathsf{GTs}}_{a})'(x)= \frac{a \; e^{ax}}{e^{ax}-1}
\]
which is defined on $(0,1]$ and $\neq 0$. Thus, 
({\bf I3}) is verified.

Finally, we easily see that:
\[
\lim_{ a \rightarrow 0}(I^{\mathsf{GTs}}_{a})'(x)= \frac{1}{x}
\]
thus ({\bf I4}) is verified.

\subsection{Schur-concavity}
\label{subSchur2}

Let us denote $h: [0,1] \rightarrow [0, \infty)$, $x \mapsto 
\frac{e^{ax}-1}{e^{a}-1}$. Then, we have:

\[
\frac{h(x)}{e^{ax}} \; \left(W^{-1} \circ I^{\mathsf{GTs}}_{a} \right)''(x)=
\frac{a^{2}}{e^{a}-1} \; u(x),
\]
where $(\cdot)''$ denotes the second derivative and 
\[
u(x)= h(x) \; \mathsf{ln}(h(x)) + 2 h(x) + \frac{1}{e^{a}-1}.
\]
The derivative of $u$ is then:
\[
u'(x)= h'(x) \; (\mathsf{ln}(h(x)) +3).
\]
Noticing that $h' > 0$ on $[0,1]$ the function $u$ is such that 
$u(0)= \frac{1}{e^{a}-1} > 0$, $u(1)= 2 + \frac{1}{e^{a}-1} > 0$ and 
admits a minimum at 
$x^{*}= \mathsf{ln}(1+ (e^{a}-1) \; e^{-3})^{\frac{1}{a}}$ which is 
$u(x^{*})=-e^{-3} + \frac{1}{e^{a}-1}$. Thus, $(W^{-1} \circ I^{\mathsf{GTs}}_{a})''
 \geq 0$ if and only if $u(x^{*}) \geq 0$ that is: 
\[
a \leq \mathsf{ln}(1+ e^{3}).
\]

\subsection{Microcanonical ensemble} 
Applying formula (\ref{eqMicro}) with (\ref{InvGTs}) we have:

\[
\forall i, \; p_{i} = \frac{1}{a} \mathsf{ln} \left(1 + (e^{a}-1) \; 
\frac{\langle 1 \rangle_{a}}{n} \right).
\]
And by (\ref{eqMicroSa}) we have:
\[
S_{a} = \langle 1 \rangle_{a} \; \mathsf{ln}(n).
\]

Note that the $\rightarrow 0$ limit of $p_{i}$ is $\frac{1}{n}$.

\subsection{Canonical ensemble}
Applying formula (\ref{eqCanon2}) with (\ref{InvGTs}) we have:
\[
\forall i, \; p_{i} = \frac{1}{a} \mathsf{ln} \left(1 + (e^{a}-1) \; 
\frac{e^{\mu_{a} \varepsilon_{i}}}{Z(\mu_{a})} \right)
\]

\subsection{Deformed algebra}

Following subsection~\ref{subGuideline} and noticing that $I^{-
  \mathsf{GTs}}_{a} \circ \mathsf{ln}(x)= \frac{1}{a} \mathsf{ln}(1 +
(e^{a}-1) \; x)$, $e^{I^{\mathsf{GTs}}_{a}(x)}=\frac{e^{ax}-1}{e^{a}-1}$
we define the deformed addition:

\begin{equation}
\label{addition1}
x \stackrel{a}{\oplus} y := \frac{1}{a} \mathsf{ln}(e^{a x} + e^{a y} -1)
\end{equation}
the deformed multiplication which is a particular case 
of copula (see e.g. \cite[Table 2.6]{kn:Alsinaetal06}):

\begin{equation}
\label{multi1}
x \stackrel{a}{\otimes}y:= \frac{1}{a} \mathsf{ln} \left(1 + \frac{(e^{a x} -1) \; (e^{a y} -1)}{e^{a}-1} \right)
\end{equation}
the deformed substraction:

\begin{equation}
\label{substraction1}
x \stackrel{a}{\ominus} y := \frac{1}{a} \mathsf{ln}(1 + e^{a x} - e^{a y})
\end{equation}
the deformed division:

\begin{equation}
\label{division1}
x \stackrel{a}{\oslash} y := \frac{1}{a} \mathsf{ln} \left(1 + (e^{a}-1) \; 
\frac{e^{a x}-1}{e^{a y}-1} \right)
\end{equation}
And the deformed calculus is as follows. We define the 
deformed derivatives by:

\begin{equation}
\label{derivee1}
D_{a}f(x) = \frac{1}{a} \mathsf{ln} \left(1 + (e^{a}-1) \; 
\frac{e^{a f(x)}}{e^{a x}} \; f'(x) \right)
\end{equation}

And the deformed integral by:
\begin{equation}
\label{integrale1}
\int^{a} f = \frac{1}{a} \mathsf{ln} \left(1+ \frac{a}{e^{a}-1} \; 
\int e^{ax} (e^{a f(x)}-1) dx \right)
\end{equation}
And we see that: 

\[
\bar{ll}
\stackrel{0}{\oplus}  & = +, \\
\stackrel{0}{\otimes}  & = \times , \\ 
\stackrel{0}{\ominus}  & = - , \\ 
\stackrel{0}{\oslash}  & = \slash , \\ 
D_{0} f                 & = f', \\
\int^{0} f & = \int f.
\ear
\]

\section{Conclusion}
\label{secConcl}

To conclude let us recall the main features of this paper. Based on
the remark that Shannon entropy can be expressed as a function of the
self-information (i.e. the logarithm) and the inverse of the Lambert $W$
function a new definition of generalized entropy has been proposed which
limit is the classical Shannon entropy.

Axioms that characterize one-parameter deformation of the logarithmic
and exponential functions were proposed (see ({\bf I1})-({\bf I3})). A
notion of one-parameter deformed probability and independence of
systems were also introduced. By using a standard variational
principle microcanonical and canonical distributions have been
established. The new notion of independence leads to define a
non-commutative and non-associative composition law which limit 
is the ussual addition.

The generalized entropy proposed in this paper allows us to define a
generalized temperature based on a fundamental relation of the
standard thermodynamics. It was also possible to define a
generalized free energy. All these quantities converge to the standard
ones.

A systematic way to define a 'deformed algebra' and 
a 'deformed calculus' was provided. The deformed algebra and calculus 
converge to the standard ones. Some of the operations defined in 
Section~\ref{secEx} are knonw as triangular norms or copulas 
(see e.g. \cite{kn:Klement00}). They appear 
in other contexts such as fuzzy logic and statistics. For example 
the multiplication defined by (\ref{multi1}) is a triangular 
norm on $[0,1]$. Note that this multiplication is also known as 
Frank copula (see e.g. \cite{kn:Nelsen06}). With $0$ 
interpreted as falsity and $1$ as truth a triangular norm is 
an operation which is commutative, associative, 
monotone, with neutral element $1$ and such that 
$0 \stackrel{a}{\otimes} x=0$ and continuous. 
A t-norm is interpreted in fuzzy logic as a conjunction. The residuum of 
$\stackrel{a}{\otimes}$ defined by (\ref{multi1}) which is 
denoted $\stackrel{a}{\Rightarrow}$ and defined as:

\[
x \stackrel{a}{\Rightarrow} y:= 
\sup \left\{ z \in [0,1]: z \stackrel{a}{\otimes} x \leq y \right\} =\min(1, y \stackrel{a}{\oslash} x)
\]
where $\stackrel{a}{\oslash}$ is the division defined by (\ref{division1}). 
Thus, roughly speaking the 
division plays the role of the implication in fuzzy logic. Using the 
standard negation $\sim : x \mapsto 1 -x$ one obtains a disjunction 
$\stackrel{a}{\mathsf{OR}}$ as follows:

\[
x \stackrel{a}{\mathsf{OR}} y:= 1 - (1 -x)\stackrel{a}{\otimes} (1-y). 
\]
Thus, interpreting $\leq$ as the deduction operator the 
structure $([0,1], \stackrel{a}{\otimes}, \sim, \stackrel{a}{\mathsf{OR}}, 
\stackrel{a}{\Rightarrow}, 0, 1, \leq)$ plays the same role 
in fuzzy logic than the Boolean algebra in classical logic. 

Finally, let us make the following remarks. (A) The $a$-independence of 
two systems (\ref{eqa-independence}) is equivalent to:
\[
p^{\Omega \times \Omega'}_{ij} = p^{\Omega}_{i} \stackrel{a}{\otimes} p^{\Omega'}_{j}=
I_{a}^{-1}(I_{a}( p^{\Omega}_{i})+ I_{a}(p^{\Omega'}_{j}))
\]
by definition of $[ \cdot ]_{a}$ (see Definition~\ref{mainDef}) and 
$\stackrel{a}{\otimes}$ (see \ref{eqGenMult}). (B) By construction 
$\stackrel{a}{\otimes}$ is a t-norm. And t-norms are also used to 
express the independence of events in the context of fuzzyness \cite[Definition 2.1]{kn:cooman97}. Thus, starting from an a-priori given notion of 
independence modelled by a t-norm $T$ it is possible to derive a deformed 
logarithm $I_{a}$ by solving $x \stackrel{a}{\otimes} y =T(x,y)$ which 
is equivalent to solve the functional equation:
\begin{equation}
\label{eqfonc}
I_{a}^{-1}(I_{a}(x)+I_{a}(y)) = T(x,y).
\end{equation}
Thus, the deformed logarithm $I_{a}$ is interpreted as an additive 
generator of the t-norm $T$. 
As a basic example let us define $T(x,y)=x y$. Then, (\ref{eqfonc}) becomes:
\[
I_{a}(x)+I_{a}(y) = I_{a}(xy)
\]
which solution is $I_{a}(x):= a \mathsf{ln}(x)$ with $a \geq 0$. This leads 
to the deformed addition being defined by: $x \stackrel{a}{\oplus} y :=
(x^{a}+y^{a})^{\frac{1}{a}}$ and the deformed entropy defined by:
\[
S_{a}:= - \sum_{i} a \; \mathsf{ln}(p_{i}) \; p_{i}^{a}.
\]

Now, looking back the interpration of an $a$-probability vector in the 
context of incomplete statistics (see Remark~\ref{remark1} in the 
introduction) the last 
point of this conclusion suggests another point of view in term of 
fuzzyness vs randomness for deformed statistics and thermodynamics.

\section*{Acknowledgement}
Author would like to thank referees for their comments, suggestions 
and remarks.

\bibliographystyle{plain}
\bibliography{ref_lt}   

\end{document}